\documentclass{article}

\PassOptionsToPackage{numbers, compress}{natbib}
\usepackage[dblblindworkshop, final]{neurips_2026}
\workshoptitle{AI Foundations for Power Grids}

\usepackage[utf8]{inputenc} 
\usepackage[T1]{fontenc}    
\usepackage{hyperref}       
\usepackage{url}            
\usepackage{booktabs}       
\usepackage{amsfonts}       
\usepackage{nicefrac}       
\usepackage{microtype}      
\usepackage{xcolor}         

\usepackage{graphicx}
\usepackage{amsmath}
\usepackage{amssymb}
\usepackage{amsthm}
\usepackage{tikz}
\usetikzlibrary{positioning,fit}
\DeclareMathOperator*{\argmin}{arg\,min}
\DeclareMathOperator*{\argmax}{arg\,max}
\newtheorem{proposition}{Proposition}
\graphicspath{{img/}}

\title{Reconciling European Electricity Data with Diffusion Priors Learned from Observations Alone}

\author{%
  Matthias Pirlet\textsuperscript{a} \quad Adrien Bolland\textsuperscript{b} \quad
  Gilles Louppe\textsuperscript{a} \quad Damien Ernst\textsuperscript{a}\\
  \textsuperscript{a}Montefiore Institute, University of Liège, Belgium\\
  \textsuperscript{b}CMAP, École Polytechnique, France\\
  \texttt{matthias.pirlet@uliege.be}
}

\begin{document}

\maketitle

\begin{abstract}
The ENTSO-E Transparency Platform publishes generation, load, and cross-border power flows for European bidding zones. The values carry undocumented errors and violate energy conservation, and data reconciliation consists in correcting them to satisfy that balance. In this work, we formulate data reconciliation as Bayesian inference. Classical methods implicitly take a uniform prior, which we replace with a diffusion model estimated from the observations alone by empirical-Bayes expectation-maximization. The resulting posterior is sampled with a guided process whose projection step makes every sample satisfy the balance exactly. On synthetic systems of six to thirty-five bidding zones, built from ENTSO-E resource-adequacy data, the diffusion prior gives the most accurate reconciliation, increasingly so as the system grows. Its credible intervals are calibrated or conservative, and it remains the most accurate under a misspecified noise covariance. Finally, when the generation of entire zones goes unreported, the classical methods fail while the learned prior recovers it from cross-zone structure.
\end{abstract}

\section{Introduction} \label{sec:introduction}

Monitoring, planning, and operating electricity systems requires accurate data on generation, load, and cross-border power flows, a need the European Union answered by requiring their publication through the ENTSO-E Transparency Platform, established in 2015 under Regulation (EU) No~543/2013~\cite{eu543-2013}. The platform reports for every bidding zone the generation by fuel type, the electrical load, and the physical flow on each border. The values are communicated by the transmission system operators (TSOs), who meter large units and interconnections, while the load is derived from generation and exchanges~\cite{eu543-2013}. The quality of the values varies across TSOs and reported quantities~\cite{hirth2018entso, ember2024, entsoe-etug}, and no TSO documents the uncertainty of the submitted data. In practice, we observe that, at any hour, the reported generation and imports of a zone do not balance its reported load and exports, with residuals sometimes reaching thousands of megawatts. We call these reported values observations. The state is the same quantities free of reporting errors and satisfying energy conservation.

Recovering the state from the observations is a data reconciliation problem, developed in process engineering for material and energy balances~\cite{narasimhan1999data}. It differs from power system state estimation, which recovers the bus voltages and angles from grid measurements, where the physics enters through the measurement equations~\cite{schweppe1970power}. Here, the state and the observations are vectors of the same quantities, and the physics enters as the zonal balance constraint. Earlier work has trained neural networks to estimate the state of transmission and distribution grids~\cite{mestav2019bayesian, he2020power} and used diffusion models to perform data reconciliation on corrupted grid measurements~\cite{pei2024detection} and dynamic state estimation~\cite{pei2025deep}, but none of it addresses bidding-zone observations.

Let us consider discrete times $t = 1, \dots, T$, over which the state and the observations evolve. We represent the bidding zones and their interconnections as a directed graph $\mathcal{G} = (\mathcal{V}, \mathcal{E})$, where the vertices $\mathcal{V}$ are the zones and the edges $\mathcal{E} \subseteq \mathcal{V} \times \mathcal{V}$ are the interconnections, each contributing the two edges $(i,j)$ and $(j,i)$, one per flow direction. Each zone $i \in \mathcal{V}$ hosts a set of generation sources $\mathcal{S}_i$. At time $t$, zone $i$ produces $g_{i,s,t}$ from each source $s \in \mathcal{S}_i$, consumes the load $d_{i,t}$, and exchanges the flow $f_{ij,t}$ on each edge $(i,j) \in \mathcal{E}$. Together these quantities compose the state $\mathbf{x}_t \in \mathbb{R}^n$, with $n = \sum_{i \in \mathcal{V}} |\mathcal{S}_i| + |\mathcal{V}| + |\mathcal{E}|$, and we call its components signals. Energy conservation, i.e., generation plus imports equal load plus exports in each zone, is expressed by the $|\mathcal{V}|$ linear constraints $\mathbf{C}\mathbf{x}_t = \mathbf{0}$ at each time, with $\mathbf{C} \in \mathbb{R}^{|\mathcal{V}| \times n}$ independent of $t$. Generation and flows are non-negative, while load can be negative, because the reported load is consumption minus local generation, and these constraints make the feasible states a polyhedral cone $\mathcal{X} \subseteq \mathbb{R}^n$. An observation $\mathbf{y}_t \in \mathbb{R}^n$ is a noisy view of the state and may violate the energy balance.

The goal is to infer, at each time $t$, the state that generated the observations, together with the remaining uncertainty. We formulate reconciliation as Bayesian inference, under the latent variable model of Figure~\ref{fig:lvm} in Appendix~\ref{app:posterior-derivations}, defined by a prior and a likelihood. For the likelihood, we model the observations as
\begin{equation} \label{eq:observation}
\mathbf{y}_t = \mathbf{A}_t (\mathbf{x}_t + \boldsymbol{\varepsilon}_t), \quad \boldsymbol{\varepsilon}_t \sim \mathcal{N}(\mathbf{0}, \boldsymbol{\Sigma}),
\end{equation}
where $\mathbf{A}_t \in \{0,1\}^{n \times n}$ is a diagonal matrix whose $k$-th diagonal entry is $1$ if signal $k$ is reported at time $t$. The observation noise $\boldsymbol{\varepsilon}_t$ is assumed to have a diagonal covariance $\boldsymbol{\Sigma}$, the same for every time $t$, so the noise is independent across signals. Under the model, $\mathbf{y}_t$ is Gaussian with mean $\mathbf{A}_t \mathbf{x}_t$, which defines the likelihood $p(\mathbf{y}_t \mid \mathbf{x}_t)$. Its only parameter is $\boldsymbol{\Sigma}$, which no TSO publicly documents, so it must be set from domain knowledge. The states themselves follow an unknown distribution $p(\mathbf{x})$ supported on $\mathcal{X}$. This source distribution is the ideal prior, and the methods differ in the prior they adopt in its place. We model the states as independent and identically distributed across time, ignoring their temporal correlation, and Bayes' theorem gives the posterior
\begin{equation} \label{eq:posterior}
p(\mathbf{x}_t \mid \mathbf{y}_t) \propto p(\mathbf{y}_t \mid \mathbf{x}_t)\, p(\mathbf{x}_t),
\end{equation}
the distribution of the state given the observation. Classical reconciliation implicitly takes the prior uniform over $\mathcal{X}$, and we replace it with a diffusion model estimated from the observations alone by empirical-Bayes expectation-maximization~\cite{rozet2024learning}.

\section{Method} \label{sec:method}

A reconciliation method fixes two ingredients, the prior and the likelihood through its covariance $\boldsymbol{\Sigma}$, and gives an algorithm for evaluating or sampling the corresponding posterior.

\paragraph{Priors.} The uniform prior over $\mathcal{X}$ is the implicit choice of classical reconciliation. In practice, however, the state distribution is far from uniform. Generation is correlated across zones, and the system moves between operating regimes, defined by the border flow directions and the units committed, so the distribution is multimodal and concentrated. A prior capturing this concentration should narrow the posterior. A Gaussian prior $\mathcal{N}(\boldsymbol{\mu}_0, \boldsymbol{\Sigma}_0)$ is unimodal, so it captures the correlations but not the regimes. A diffusion prior, a denoising network defining a distribution $p_\theta(\mathbf{x})$ over states~\cite{song2021scorebased, karras2022elucidating}, can be multimodal and represent both. Its parameterization is in Appendix~\ref{app:diffusion-details}.

\paragraph{Estimation.} The uniform prior has no free parameters. The Gaussian and diffusion priors are estimated by empirical Bayes (EB), from historical observations and without access to the states. For the Gaussian EB prior, $\boldsymbol{\mu}_0$ and $\boldsymbol{\Sigma}_0$ are estimates of the mean and covariance of the states. The states are not observed, but at times where every signal is reported, $\mathbf{y}_t = \mathbf{x}_t + \boldsymbol{\varepsilon}_t$, so the model gives the mean and covariance of the states in closed form as $\mathbb{E}[\mathbf{y}_t]$ and $\mathrm{Cov}(\mathbf{y}_t) - \boldsymbol{\Sigma}$. Replacing both by their sample versions over those times gives $\boldsymbol{\mu}_0$ and $\boldsymbol{\Sigma}_0$. The sample estimate of $\boldsymbol{\Sigma}_0$ need not be positive semidefinite, so we clip its negative eigenvalues to a small positive floor. Times with unreported signals contribute nothing to this estimator. The diffusion prior should represent the state distribution, but training its denoiser requires state samples, which are unavailable, so the expectation-maximization (EM) scheme of Rozet et al.~\cite{rozet2024learning} estimates it from the observations. The expectation step (E-step) draws plausible states for each observation by sampling the posterior under the current prior, and the maximization step (M-step) refits the network on those samples. The iterations carry the usual EM guarantees~\cite{wu1983convergence, mclachlan2008algorithm, balakrishnan2017statistical}. Only the network is estimated, while the covariance $\boldsymbol{\Sigma}$ keeps the value specified from domain knowledge and enters only the E-step posterior. We call the result the diffusion EB prior. Appendix~\ref{app:diffusion-details} gives the training loss, the EM hyperparameters, and the guided posterior sampler. The oracle prior, an upper bound, trains the same model on the states, which real data never provides.

\paragraph{Inference.} Under the uniform prior, classical reconciliation returns the maximum-a-posteriori (MAP) estimate. This is the weighted least squares (WLS) estimator, which solves the quadratic program
\begin{equation} \label{eq:wls-qp}
\begin{aligned}
\hat{\mathbf{x}}_t = \argmax_{\mathbf{x}} \; p(\mathbf{x} \mid \mathbf{y}_t) &= \argmin_{\mathbf{x}} \; (\mathbf{y}_t - \mathbf{A}_t \mathbf{x})^\top \boldsymbol{\Sigma}^{-1} (\mathbf{y}_t - \mathbf{A}_t \mathbf{x}) \\
&\qquad \text{s.t.} \quad \mathbf{C}\mathbf{x} = \mathbf{0}, \; \mathbf{g} \geq \mathbf{0}, \; \mathbf{f} \geq \mathbf{0},
\end{aligned}
\end{equation}
where $\mathbf{g}$ and $\mathbf{f}$ stack the generation and flow signals of $\mathbf{x}$. The objective is the negative log-likelihood, and the constraints restrict $\mathbf{x}$ to $\mathcal{X}$, the support of the prior. The WLS variants compared in Section~\ref{sec:experiments} differ from~\eqref{eq:wls-qp} in the covariance they use in place of the unknown $\boldsymbol{\Sigma}$ and, for naive and oracle WLS, in dropping the non-negativity constraints to obtain a closed form. Under the Gaussian EB prior the posterior is Gaussian in closed form, conditioned on the balance (Appendix~\ref{app:posterior-derivations}). Under the diffusion prior it has no closed form and must be sampled. Each step of the guided denoising process~\cite{song2021denoising} is pulled toward the observation through the likelihood and orthogonally projected onto the null space of $\mathbf{C}$ as in DDNM~\cite{wang2023zeroshot}, so every sample satisfies the balance exactly (Appendix~\ref{app:diffusion-details}). The projection enforces only the equality constraint $\mathbf{C}\mathbf{x} = \mathbf{0}$, while the non-negativity inequalities have no closed-form projection, so posterior draws can violate them. We call the posterior under the diffusion prior the diffusion posterior, and an ensemble of draws gives its credible intervals.

\section{Experiments} \label{sec:experiments}

Reconciliation accuracy cannot be measured on real ENTSO-E data, which lacks known states. Instead we build four synthetic systems of increasing size from the ENTSO-E European Resource Adequacy Assessment (ERAA) 2024~\cite{eraa2024}, each a set of bidding zones and their interconnections, named \textsc{eu6} (6 zones, $n = 108$ signals), \textsc{euwest}, \textsc{euwide}, and \textsc{eufull} (35 zones, $n = 651$). For each of the 35 climate years, the weather scenarios of ERAA, an hourly merit-order dispatch produces states $\mathbf{x}_t$ satisfying the balance exactly, and the observations $\mathbf{y}_t$ are these states plus independent Gaussian noise with per-fuel, per-zone standard deviations between 50 and 300~MW. The learned priors are trained on the observations of climate years 1-28, never on the states, and every method is evaluated on years 32-35 ($35\,040$ hours) against the states. Appendix~\ref{app:exp-details} details the systems, the noise model, and the de Chalendar heuristic, and Table~\ref{tab:methods} summarizes the compared methods.

We compare naive WLS ($\boldsymbol{\Sigma} = \mathbf{I}$), oracle WLS given the true $\boldsymbol{\Sigma}$, the constrained WLS QP adding non-negativity, the heuristic of de Chalendar and Benson~\cite{de2021physics}, the Gaussian EB prior, and the diffusion EB prior. RMSE against the states scores each method's best point estimate, the posterior mean for the learned priors, and the unreconciled observations give the reference error. The posteriors are judged separately, by the coverage of their credible intervals and by the CRPS.
\begin{table}[t]
\centering
\caption{Overall RMSE (MW) and CRPS (MW) of the three posteriors on the test set, and RMSE (MW) on the masked signals when the generation of the six core zones is masked.}
\label{tab:main-results}
\renewcommand{\arraystretch}{1.1}
\setlength{\tabcolsep}{2pt}\scriptsize
\begin{tabular}{lrrrrrrrrrrrr}
\toprule
 & \multicolumn{4}{c}{RMSE, fully observed} & \multicolumn{4}{c}{CRPS, fully observed} & \multicolumn{4}{c}{RMSE on masked signals} \\
\cmidrule(lr){2-5}\cmidrule(lr){6-9}\cmidrule(lr){10-13}
Method & \textsc{eu6} & \textsc{euwest} & \textsc{euwide} & \textsc{eufull} & \textsc{eu6} & \textsc{euwest} & \textsc{euwide} & \textsc{eufull} & \textsc{eu6} & \textsc{euwest} & \textsc{euwide} & \textsc{eufull} \\
\midrule
Raw observations   & 146.4 & 142.4 & 138.2 & 140.8 & -- & -- & -- & -- & -- & -- & -- & -- \\
Naive WLS          & 141.9 & 138.0 & 134.0 & 136.5 & -- & -- & -- & -- & -- & -- & -- & -- \\
Oracle WLS         & 134.9 & 130.1 & 126.9 & 129.3 & 75.4 & 79.3 & 84.8 & 84.7 & -- & -- & -- & -- \\
Constrained WLS QP & 115.1 & 110.0 & 104.0 & 103.9 & -- & -- & -- & -- & 5\,197 & 5\,311 & 5\,580 & 5\,615 \\
De Chalendar~\cite{de2021physics} & 287.3 & 232.7 & 197.9 & 194.2 & -- & -- & -- & -- & -- & -- & -- & -- \\
Gaussian EB prior (ours) & 117.6 & 106.1 & 90.6 & 87.6 & 61.6 & 60.3 & 54.6 & 51.1 & 2\,050 & 1\,647 & 1\,401 & 1\,255 \\
\textbf{Diffusion EB prior (ours)} & \textbf{99.8} & \textbf{92.0} & \textbf{80.5} & \textbf{79.0} & \textbf{52.0} & \textbf{52.7} & \textbf{51.7} & \textbf{50.1} & \textbf{1\,920} & \textbf{1\,411} & \textbf{1\,149} & \textbf{1\,035} \\
\bottomrule
\end{tabular}
\end{table}

The diffusion EB prior is the most accurate on every system, as the left half of Table~\ref{tab:main-results} shows, and the RMSE gap to the constrained WLS QP widens from $13.3\%$ to $24.0\%$ as the system grows. The Gaussian EB prior, which captures the correlations but not the regimes, splits this gain in two, the correlations driving the growth with system size and the regimes contributing the rest. Naive and oracle WLS improve on the raw observations in proportion to what they know of $\boldsymbol{\Sigma}$, and the de Chalendar heuristic is the only method worse than the raw observations, because its magnitude-based weights mismatch our per-fuel noise and its under-weighted loads absorb the balance violations. Table~\ref{tab:base-results} shows that the advantage of the diffusion EB prior holds on every component.

A posterior is useful only if its credible intervals can be trusted. It is calibrated when the intervals contain the state as often as they claim. The diffusion posterior is calibrated on the two smaller systems, at worst 88.6\% coverage for a 90\% target, and conservative on the larger ones, where the intervals contain the state more often than they claim, as shown in Figure~\ref{fig:calibration}. The continuous ranked probability score (CRPS), the standard score for probabilistic forecasts, is lower when the posterior is calibrated and its intervals are narrow, and the diffusion posterior has the best CRPS on every system, as Table~\ref{tab:main-results} shows. The uniform prior also yields a full posterior, the balance-conditioned Gaussian whose mode is the WLS estimate (Appendix~\ref{app:posterior-derivations}), with wider intervals than the diffusion posterior.

\begin{figure*}[t]
\centering
\includegraphics[width=0.85\linewidth]{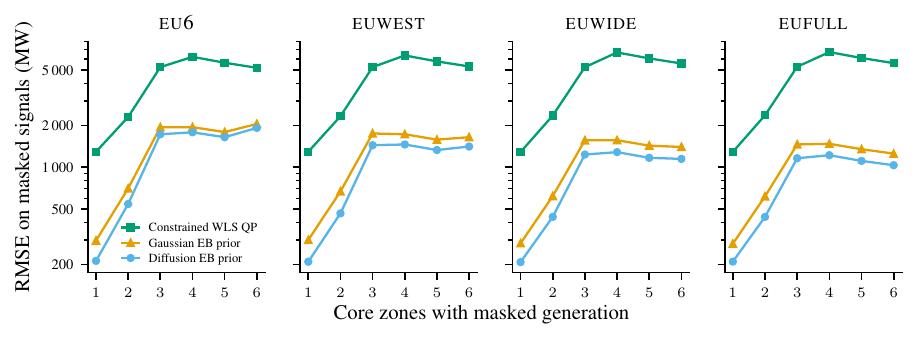}
\caption{RMSE on the masked signals against the number of core zones with masked generation.}
\label{fig:masking-curves}
\end{figure*}

Masking the generation of the six core zones, while their load and border flows stay reported, tests recovery of unreported generation. The same 84 signals are masked on every system, so an overall error would shrink with system size by averaging alone, and we score only the masked signals. For a masked zone, the balance fixes the total generation, which all methods recover within 360-610~MW, but not its division among the fuels. A uniform prior carries no information on that division, so the constrained WLS QP fails, its error exceeding $5\,000$~MW on every system. The learned priors infer the division from the zones that still report, more accurately as the system grows, as Figure~\ref{fig:masking-curves} shows, and the diffusion EB prior's advantage over the Gaussian one grows from 6\% to 18\%.

The covariance $\boldsymbol{\Sigma}$ is a belief, so we re-estimate and evaluate both learned priors with it scaled by 0.5 and by 2. WLS is immune to this stress, since a global rescaling of $\boldsymbol{\Sigma}$ leaves its minimizer unchanged, and the two learned priors fail in opposite directions. Overstating the noise costs the diffusion prior little and understating erases its advantage, while the Gaussian EB prior tolerates understating but degrades past the raw observations when the noise is overstated, since its covariance subtraction then removes variance the states actually have. The diffusion prior stays the most accurate under every belief. The diffusion EB prior is not a faithful recovery of the state distribution, its marginals $77\%$ too wide, yet it reconciles best, because reconciliation is governed by the likelihood and a wide, well-correlated prior still regularizes the posterior. The prior predictive comparison of Appendix~\ref{ssec:prior-fidelity}, prior samples plus the believed noise against the observed $\mathbf{y}_t$, reproduces their spread and balance-gap magnitude, a ground-truth-free check that carries over to real data. Appendix~\ref{app:additional-results} gives the complete results.

\section{Conclusion} \label{sec:conclusion}

We formulated ENTSO-E data reconciliation as Bayesian inference and replaced the uniform prior of classical reconciliation with a diffusion model estimated from observations alone. On synthetic systems with known ground truth, the diffusion prior gives the most accurate reconciliation with calibrated or conservative credible intervals, and it recovers unreported generation where the uniform prior fails. The observation noise of the benchmark remains Gaussian, independent, and constant per fuel, while real reporting errors scale with magnitude, are autocorrelated, and carry per-TSO bias that the empirical-Bayes loop cannot remove~\cite{kazmi2022good}, though a magnitude-dependent $\boldsymbol{\Sigma}$ fits the same likelihood. Posterior draws do not enforce non-negativity, which predict-project-renoise~\cite{rochman2026predict} would impose. On real data, validation rests on balance residuals and cross-source references.

\begin{ack}
The authors would like to thank Engie GMA's team, especially Alexandre Huynen, for sharing expert knowledge about the problem. Adrien Bolland acknowledges the financial support received during part of this research through a research fellowship from the F.R.S.-FNRS.
\end{ack}

\section*{Use of LLMs.} We used an LLM for grammar and spelling correction and sometimes to suggest clearer phrasing. All research content, including the method, the experiments, the results and their interpretation, is our own. We reviewed every sentence and take full responsibility for the content of the paper.

\bibliographystyle{unsrtnat}
\bibliography{bibliography}

\appendix

\section{Posterior under a Gaussian likelihood: uniform vs.\ Gaussian prior}
\label{app:posterior-derivations}

This appendix derives the Gaussian posteriors under the uniform prior of the WLS family and under a generic Gaussian prior, which includes the Gaussian EB prior of Section~\ref{sec:method}. Proposition~\ref{prop:flat} gives the posterior under the uniform prior, Proposition~\ref{prop:gaussian-prior} under a Gaussian prior. The last subsection conditions both on the balance $\mathbf{C}\mathbf{x} = \mathbf{0}$. The experiments sample the Gaussian EB and oracle WLS posteriors from this conditioned form.

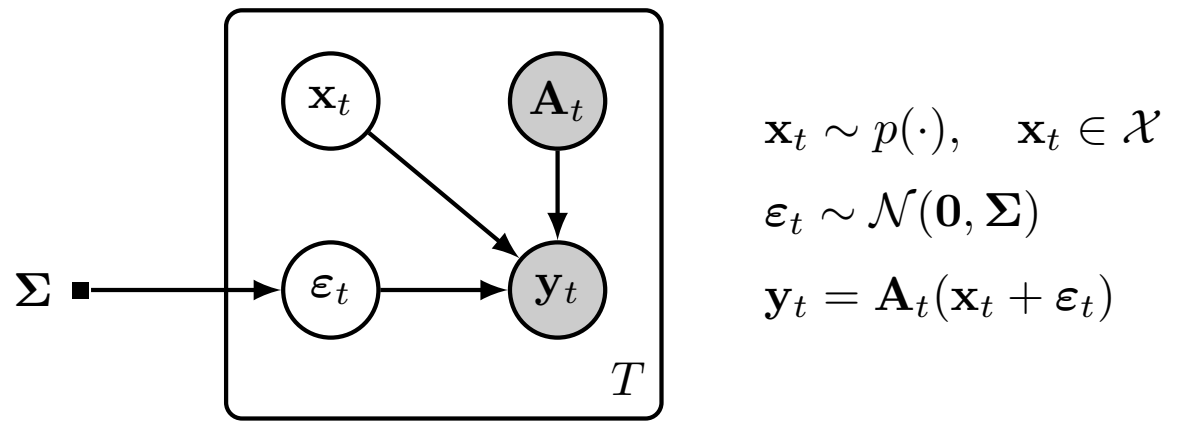
\begin{figure}[h]
\centering
\begin{tikzpicture}[>=latex, thick, auto, scale=1, every node/.style={scale=0.9}]
\tikzstyle{var}=[circle, draw=black, minimum size=0.7cm, inner sep=0pt]
\tikzstyle{obs}=[circle, draw=black, fill=gray!40, minimum size=0.7cm, inner sep=0pt]
\tikzstyle{box}=[rectangle, draw=black, rounded corners=3pt, inner sep=5pt]
\node[box, minimum width=3.2cm, minimum height=3.0cm] (plate) at (0,0) {};
\node[var] (x)   at ([xshift=-2.2cm, yshift=0.75cm]plate.east) {$\mathbf{x}_t$};
\node[var] (eps) at ([xshift=-2.2cm, yshift=-0.5cm]plate.east) {$\boldsymbol{\varepsilon}_t$};
\node[obs] (A)   at ([xshift=-0.7cm, yshift=0.75cm]plate.east) {$\mathbf{A}_t$};
\node[obs] (y)   at ([xshift=-0.7cm, yshift=-0.5cm]plate.east) {$\mathbf{y}_t$};
\node[rectangle, fill=black, minimum size=0.12cm, inner sep=0pt,
      label={left:$\boldsymbol{\Sigma}$}] (Sig) at (-2.4cm,-0.5cm) {};
\node[anchor=north east] at ([yshift=0.5cm]plate.south east) {$T$};
\draw[->] (x) -- (y);
\draw[->] (eps) -- (y);
\draw[->] (A) -- (y);
\draw[->] (Sig) -- (eps);
\node[anchor=west] at (2.0cm,  0.55cm) {$\mathbf{x}_t \sim p(\cdot), \quad \mathbf{x}_t \in \mathcal{X}$};
\node[anchor=west] at (2.0cm,  0.0cm)  {$\boldsymbol{\varepsilon}_t \sim \mathcal{N}(\mathbf{0}, \boldsymbol{\Sigma})$};
\node[anchor=west] at (2.0cm, -0.55cm) {$\mathbf{y}_t = \mathbf{A}_t(\mathbf{x}_t + \boldsymbol{\varepsilon}_t)$};
\end{tikzpicture}
\caption{The latent variable model for data reconciliation. White circles are
latent variables, gray circles are observed ones, and the black square is a
fixed parameter. The rectangular box denotes the $T$ times, modeled independently, and
$\boldsymbol{\Sigma}$ sits outside it, shared by all times. The mask $\mathbf{A}_t$ is read from which signals
the platform publishes at time $t$, and the observation $\mathbf{y}_t$ is a
deterministic function of its three parents through~\eqref{eq:observation}.}
\label{fig:lvm}
\end{figure}

Let $\mathbf{x} \in \mathbb{R}^n$ be the latent state and $\mathbf{y} \in \mathbb{R}^n$ the observation. We drop the time subscript $t$. The likelihood is linear-Gaussian,
\begin{equation}
p(\mathbf{y} \mid \mathbf{x})
  = \mathcal{N}(\mathbf{y};\, \mathbf{A}\mathbf{x},\, \boldsymbol{\Sigma}),
\end{equation}
where $\mathbf{A}$ is the diagonal mask of \eqref{eq:observation} and $\boldsymbol{\Sigma} \succ \mathbf{0}$ is the noise covariance of the main text. As a function of $\mathbf{x}$, this density is proportional to the likelihood for any mask, which is all Bayes' theorem uses. Since $\boldsymbol{\Sigma}$ is diagonal, its exponent is a sum of one residual per signal, $(y_k - (\mathbf{A}\mathbf{x})_k)^2 / \Sigma_{kk}$, and an unreported signal has $y_k = 0$ and $(\mathbf{A}\mathbf{x})_k = 0$, so its residual is zero and only the reported signals enter.

\subsection{Uniform prior}

The main text takes the prior uniform over $\mathcal{X}$. To keep the posterior Gaussian, this subsection relaxes it to the prior uniform over $\mathbb{R}^n$, and the conditioning of the last subsection restores the balance exactly. Only the non-negativity constraints of $\mathcal{X}$ are then missing, so posterior draws can violate them (Section~\ref{sec:conclusion}). Either uniform prior is improper, $\mathcal{X}$ being an unbounded cone, but the posterior below is proper when every signal is reported.

\begin{proposition} \label{prop:flat}
Under the uniform prior $p(\mathbf{x}) \propto 1$, with $\mathbf{A}$ of full column rank, the posterior is Gaussian,
\begin{equation} \label{eq:posterior-flat}
p(\mathbf{x} \mid \mathbf{y})
  = \mathcal{N}\!\left(\mathbf{x};\,
      \boldsymbol{\mu}_{\mathbf{x}\mid\mathbf{y}},\,
      \boldsymbol{\Sigma}_{\mathbf{x}\mid\mathbf{y}}\right),
\end{equation}
with
\begin{equation} \label{eq:flat-params}
\boldsymbol{\Sigma}_{\mathbf{x}\mid\mathbf{y}}
  = \bigl(\mathbf{A}^\top \boldsymbol{\Sigma}^{-1} \mathbf{A}\bigr)^{-1},
\qquad
\boldsymbol{\mu}_{\mathbf{x}\mid\mathbf{y}}
  = \boldsymbol{\Sigma}_{\mathbf{x}\mid\mathbf{y}}\,
    \mathbf{A}^\top \boldsymbol{\Sigma}^{-1} \mathbf{y}.
\end{equation}
Its mean is the minimizer of the unconstrained objective of \eqref{eq:wls-qp}.
\end{proposition}

\begin{proof}
With a uniform prior the log-posterior equals the log-likelihood up to an additive constant, a quadratic in $\mathbf{x}$ with Hessian $\mathbf{A}^\top \boldsymbol{\Sigma}^{-1} \mathbf{A} \succ \mathbf{0}$ by the full-rank assumption. Completing the square gives \eqref{eq:posterior-flat} with parameters \eqref{eq:flat-params}.
\end{proof}

The full column rank required here means, for the diagonal mask, that every signal is reported. When a signal is unreported, it drops out of the likelihood, so the observations carry no information about it and the inverse in \eqref{eq:flat-params} does not exist. The posterior has infinite variance along the unreported signals, and is then called improper. The missing information must come from elsewhere, from the prior, or, under the uniform prior, from the balance and the non-negativity constraints. The maximum-a-posteriori estimate is then the constrained program of \eqref{eq:wls-qp}.

\subsection{Gaussian prior}

\begin{proposition} \label{prop:gaussian-prior}
Under the Gaussian prior $p(\mathbf{x}) = \mathcal{N}(\mathbf{x};\,
\boldsymbol{\mu}_0,\, \boldsymbol{\Sigma}_0)$ with $\boldsymbol{\Sigma}_0 \succ
\mathbf{0}$, the posterior is Gaussian,
\begin{equation} \label{eq:posterior-gauss}
p(\mathbf{x} \mid \mathbf{y})
  = \mathcal{N}\!\left(\mathbf{x};\,
      \boldsymbol{\mu}_{\mathbf{x}\mid\mathbf{y}},\,
      \boldsymbol{\Sigma}_{\mathbf{x}\mid\mathbf{y}}\right),
\end{equation}
with
\begin{align}
\boldsymbol{\Sigma}_{\mathbf{x}\mid\mathbf{y}}
  &= \bigl(\mathbf{A}^\top \boldsymbol{\Sigma}^{-1} \mathbf{A}
          + \boldsymbol{\Sigma}_0^{-1}\bigr)^{-1}, \label{eq:gauss-cov} \\
\boldsymbol{\mu}_{\mathbf{x}\mid\mathbf{y}}
  &= \boldsymbol{\Sigma}_{\mathbf{x}\mid\mathbf{y}}\,
     \bigl(\mathbf{A}^\top \boldsymbol{\Sigma}^{-1} \mathbf{y}
          + \boldsymbol{\Sigma}_0^{-1} \boldsymbol{\mu}_0\bigr).
  \label{eq:gauss-mean}
\end{align}
\end{proposition}

\begin{proof}
The log-posterior is the sum of two quadratics in $\mathbf{x}$, with Hessian $\boldsymbol{\Lambda} = \mathbf{A}^\top \boldsymbol{\Sigma}^{-1} \mathbf{A} + \boldsymbol{\Sigma}_0^{-1}$, positive definite as the sum of a positive semidefinite and a positive definite matrix, and linear coefficient $\boldsymbol{\eta} = \mathbf{A}^\top \boldsymbol{\Sigma}^{-1} \mathbf{y} + \boldsymbol{\Sigma}_0^{-1} \boldsymbol{\mu}_0$. Completing the square gives a Gaussian with covariance $\boldsymbol{\Lambda}^{-1}$ and mean $\boldsymbol{\Lambda}^{-1}\boldsymbol{\eta}$, which are \eqref{eq:gauss-cov} and \eqref{eq:gauss-mean}.
\end{proof}

\subsection{Conditioning on the balance}
\label{app:balance-conditioning}

Both posteriors above have the form $\mathbf{x} \mid \mathbf{y} \sim \mathcal{N}(\mathbf{m}, \mathbf{S})$, where $\mathbf{m}$ and $\mathbf{S}$ abbreviate the mean and covariance of either proposition, and ignore the balance. Conditioning on $\mathbf{C}\mathbf{x} = \mathbf{0}$ keeps only the states that close the balance. The pair $(\mathbf{x}, \mathbf{C}\mathbf{x})$ is jointly Gaussian, so the standard Gaussian conditioning formula applies and gives the map
\begin{equation} \label{eq:balance-map}
\mathbf{M}_{\mathbf{S}} = \mathbf{I} - \mathbf{S}\mathbf{C}^\top\bigl(\mathbf{C}\mathbf{S}\mathbf{C}^\top\bigr)^{-1}\mathbf{C},
\end{equation}
\begin{equation} \label{eq:balance-moments}
\mathbf{m}_c = \mathbf{M}_{\mathbf{S}}\,\mathbf{m}, \qquad \mathbf{S}_c = \mathbf{M}_{\mathbf{S}}\,\mathbf{S},
\end{equation}
where $\mathbf{C}\mathbf{S}\mathbf{C}^\top$ is invertible because $\mathbf{S} \succ \mathbf{0}$ and $\mathbf{C}$ has full row rank. The formulas have a direct reading. The residual $\mathbf{C}\mathbf{m}$ is the balance gap of the mean, one entry per zone. The map subtracts a correction that spreads this gap over the signals according to their variances and correlations in $\mathbf{S}$, and $\mathbf{m}_c$ is the closest state to $\mathbf{m}$ that closes the balance, in the metric weighted by $\mathbf{S}^{-1}$, the reconciliation of $\mathbf{m}$ itself. The covariance $\mathbf{S}_c$ removes all variance along the $|\mathcal{V}|$ directions the balance fixes and keeps the rest, so every draw satisfies the balance exactly, $\mathbf{C}\mathbf{m}_c = \mathbf{0}$ and $\mathbf{C}\mathbf{S}_c = \mathbf{0}$, and $\mathbf{S}_c$ is singular with rank $n - |\mathcal{V}|$. Draws are therefore generated from a factorization of $\mathbf{S}_c$ rather than its inverse. Neither posterior imposes the non-negativity constraints. For $\mathbf{S} = \mathbf{I}$ the map reduces to the orthogonal projection $\mathbf{P}$ of the diffusion sampler (Appendix~\ref{app:diffusion-details}). The Gaussian EB posterior of Section~\ref{sec:method} applies the map to Proposition~\ref{prop:gaussian-prior} with $\mathbf{A} = \mathbf{I}$. The oracle WLS posterior of the calibration experiment applies it to Proposition~\ref{prop:flat}.

For the uniform prior, $\mathbf{m} = \mathbf{y}$ and $\mathbf{S} = \boldsymbol{\Sigma}$, so $\mathbf{M}_{\boldsymbol{\Sigma}}$ is the balance-only WLS reconciliation map and the conditioned posterior is $\mathcal{N}(\mathbf{M}_{\boldsymbol{\Sigma}}\mathbf{y}, \mathbf{M}_{\boldsymbol{\Sigma}}\boldsymbol{\Sigma})$. Adding fresh noise to the observation and reconciling the result, $\tilde{\mathbf{x}} = \mathbf{M}_{\boldsymbol{\Sigma}}(\mathbf{y} + \boldsymbol{\varepsilon})$ with $\boldsymbol{\varepsilon} \sim \mathcal{N}(\mathbf{0}, \boldsymbol{\Sigma})$, draws exactly from this posterior, because $\mathbf{M}_{\boldsymbol{\Sigma}}\boldsymbol{\Sigma}\mathbf{M}_{\boldsymbol{\Sigma}}^\top = \mathbf{M}_{\boldsymbol{\Sigma}}\boldsymbol{\Sigma}$. Each draw is one extra WLS solve. The identity holds for the balance constraint alone, not for the quadratic program with non-negativity.

\section{The diffusion prior}
\label{app:diffusion-details}

This appendix gives the parameterization of the diffusion prior of Section~\ref{sec:method}, its EM training, and the guided posterior sampler.

\subsection{Parameterization}

A diffusion model pairs a fixed forward process that gradually noises the state with a learned reverse process that undoes it. The forward process adds Gaussian noise to the state, producing $\mathbf{x}_\sigma \sim \mathcal{N}(\mathbf{x}_\sigma; \mathbf{x}, \sigma^2 \mathbf{I})$, with marginals ranging from the state distribution at $\sigma = 0$ to pure noise as $\sigma$ grows large~\cite{song2021scorebased}. This $\sigma$ is the noise level of the diffusion process, distinct from the observation noise $\boldsymbol{\Sigma}$ of \eqref{eq:observation}. A denoiser $d_\theta(\mathbf{x}_\sigma, \sigma)$ estimates the state from its noised version. New states are generated by reversing the process, starting from pure noise and denoising down to $\sigma = 0$, so the denoiser defines a distribution over states, the prior $p_\theta(\mathbf{x})$. Training the denoiser on states would minimize the denoising loss of Karras et al.~\cite{karras2022elucidating}
\begin{equation} \label{eq:edm-loss}
\mathcal{L}(\theta) = \mathbb{E}_{\substack{p(\mathbf{x}),\ p(\sigma),\ p(\mathbf{x}_\sigma \mid \mathbf{x})}} \bigl[ \lambda_\sigma \, \|d_\theta(\mathbf{x}_\sigma, \sigma) - \mathbf{x}\|_2^2 \bigr],
\end{equation}
where $p(\mathbf{x})$ is the state distribution, $p(\sigma)$ a distribution over noise levels, $p(\mathbf{x}_\sigma \mid \mathbf{x})$ the forward process, and $\lambda_\sigma$ a noise-dependent weighting. We have no samples of the state, only the observations $\mathbf{y}_t$, so this loss cannot be evaluated directly.

\subsection{EM training}

The EM scheme of~\cite{rozet2024learning} estimates the denoiser from the observations alone, alternating two steps over iterations $k = 1, \dots, K$. The E-step fills in plausible states for each observation, drawing $K_s$ samples from the posterior $p_{\theta_{k-1}}(\mathbf{x}_t \mid \mathbf{y}_t, \mathbf{A}_t, \boldsymbol{\Sigma})$ with the moment-matching posterior sampler (MMPS) of~\cite{rozet2024learning}. The M-step minimizes \eqref{eq:edm-loss} on those samples for $E_{\mathrm{refit}}$ epochs from a fresh optimizer. As a warm-up, our implementation first fits the denoiser to the observations for $E_{\mathrm{warm}}$ epochs, an M-step with the observations in place of posterior samples, so the initial prior still carries the observation noise, which the iterations sharpen toward the state distribution. We run $K = 8$ iterations with $K_s = 8$ posterior samples per observation, $E_{\mathrm{warm}} = 200$ warm-up epochs, and $E_{\mathrm{refit}} = 400$ refit epochs per iteration. The denoiser is a four-block residual MLP of width 256 to 1024, growing with the system, and both the E-step and inference samplers use 64 reverse steps. Reconciling one time step with 64 draws thus takes 4\,096 network evaluations, about 5~GFLOP on \textsc{eu6} and 82~GFLOP on \textsc{eufull}, and training takes about $2 \times 10^{10}$ forward-pass equivalents, dominated by the M-steps. The Gaussian EB prior uses no network and needs about $10^6$~FLOP per time step on \textsc{eufull}.

\subsection{Guided posterior sampling}

At inference we sample from the learned prior with a denoising diffusion implicit model (DDIM) reverse process~\cite{song2021denoising} on normalized states, standardized per signal with the mean and variance of the training observations. At every step, the MMPS correction replaces the denoised estimate $\hat{\mathbf{x}}_0$ (the subscript is the noise level $\sigma = 0$, not a time) by the Gaussian update $\hat{\mathbf{x}}_0 + \mathbf{V}_\sigma (\mathbf{V}_\sigma + \boldsymbol{\Sigma})^{-1} (\mathbf{y}_t - \hat{\mathbf{x}}_0)$, with $\mathbf{V}_\sigma = \sigma^2 \mathbf{I} - \sigma^4 (\boldsymbol{\Sigma}_x + \sigma^2 \mathbf{I})^{-1}$ the covariance of the state given its noised version under a Gaussian prior with a fixed covariance $\boldsymbol{\Sigma}_x$, estimated by moment matching from observations, so the update needs no differentiation of the denoiser. Signals with small observation noise are pulled strongly toward their reported values, and unreported signals, carrying a huge variance in $\boldsymbol{\Sigma}$, are left to the prior. To enforce the balance, each step then projects the corrected estimate onto the null space of $\mathbf{C}$ with $\mathbf{P} = \mathbf{I} - \mathbf{C}^\top(\mathbf{C}\mathbf{C}^\top)^{-1}\mathbf{C}$ (denoising diffusion null-space model, DDNM~\cite{wang2023zeroshot}). In image restoration DDNM also reinserts the observation into the estimate, but our constraint $\mathbf{C}\mathbf{x} = \mathbf{0}$ carries no observation, so the step is a pure projection. The projection acts in normalized coordinates, where the mean shift of the normalization leaves a small balance residual in physical units. A final projection in physical units removes it, so every sample satisfies the balance to numerical precision. The guided reverse process draws approximately from the posterior implied by the learned prior and the likelihood.

\section{Experimental details}
\label{app:exp-details}

ERAA combines the installed capacities, demand, net transfer capacities, and renewable capacity factors of a target year with 35 historical weather years, the climate years. Grounding the benchmark in this model gives it the structure of the real European system. The interconnections come from the ERAA net-transfer-capacity data, and each zone carries 14 fuel categories. Each system contains the previous one and adds zones and their interconnections:
\begin{itemize}
  \item \textbf{\textsc{eu6}} ($|\mathcal{V}| = 6$, $|\mathcal{E}| = 18$, $n = 108$): the base system $\mathrm{BE}$, $\mathrm{NL}$, $\mathrm{DE}$, $\mathrm{FR}$, $\mathrm{CH}$, $\mathrm{AT}$.
  \item \textbf{\textsc{euwest}} ($|\mathcal{V}| = 14$, $|\mathcal{E}| = 40$, $n = 250$): adds Iberia ($\mathrm{ES}$, $\mathrm{PT}$), Great Britain ($\mathrm{UK}$), and the Nordic zones $\mathrm{FI}$ and $\mathrm{SE1}$-$\mathrm{SE4}$.
  \item \textbf{\textsc{euwide}} ($|\mathcal{V}| = 28$, $|\mathcal{E}| = 102$, $n = 522$): adds Italy (seven bidding zones) and the complete Nordic set ($\mathrm{DK}$, $\mathrm{NO}$).
  \item \textbf{\textsc{eufull}} ($|\mathcal{V}| = 35$, $|\mathcal{E}| = 126$, $n = 651$): adds Central Europe ($\mathrm{CZ}$, $\mathrm{SK}$, $\mathrm{HU}$, $\mathrm{SI}$) and the Baltics ($\mathrm{EE}$, $\mathrm{LT}$, $\mathrm{LV}$).
\end{itemize}

We take 2026 as the target year, so every climate year runs the same installed capacities and network through different weather, giving 35 years of hourly data from a single unchanged system. The dispatch is a linear program and is deterministic, one realization per climate year, so the only randomness in the observations $\mathbf{y}_t$ is the observation noise. Each border flow is bounded by its net transfer capacity and Kirchhoff's laws are not imposed, so the benchmark flows are exchanges without loop flows, while the platform reports physical flows. The noise is independent and zero-mean, so $\boldsymbol{\Sigma}$ is diagonal with per-signal standard deviations
\begin{equation} \label{eq:sigma-decomp}
\sigma_i = \sigma^{\mathrm{base}}_{\mathrm{type}(i)} \cdot m_{\mathrm{zone}(i)},
\end{equation}
where $\sigma^{\mathrm{base}}_{\mathrm{type}(i)}$ is a base standard deviation for each signal type and $m_{\mathrm{zone}(i)}$ a per-zone multiplier. Each fuel has one constant base, a plausible few percent of a continental zone's typical magnitude. Onshore wind and solar are the noisiest at 300~MW, cross-border flows 200~MW, load 150~MW, and well-metered sources such as nuclear and run-of-river the cleanest at 50~MW. The per-zone multipliers lie in $[0.8, 1.2]$ and add mild TSO-to-TSO variation, for which no published ranking exists. Every signal is reported, so the mask of Eq.~\eqref{eq:observation} is the identity, and partial reporting is studied separately in the masking experiment. Holding the noise constant per fuel keeps the benchmark controlled, so the comparison isolates the reconciliation method rather than the noise model. We split along the climate-year axis to avoid temporal leakage, giving $245\,280$ training hours, $26\,280$ for validation, and $35\,040$ for testing. The states enter only the evaluation.

\begin{table}[h]
\centering
\caption{Reconciliation methods compared. ``$\Sigma$'' is the noise covariance the method uses (or its substitute). ``Historical data'' is what the method sees beyond the observations it reconciles.}
\label{tab:methods}
\renewcommand{\arraystretch}{1.15}
\begin{tabular}{lll}
\toprule
Method & $\Sigma$ & Historical data \\
\midrule
Naive WLS & $\mathbf{I}$ & None \\
Oracle WLS & True $\boldsymbol{\Sigma}$ & None \\
Constrained WLS QP & True $\boldsymbol{\Sigma}$ & None \\
De Chalendar (Eq.~\eqref{eq:dechalendar-weights}) & From magnitudes & None \\
Gaussian EB prior & True $\boldsymbol{\Sigma}$ & CY 1-28 observations \\
Diffusion EB prior~\cite{rozet2024learning} & True $\boldsymbol{\Sigma}$ & CY 1-28 observations \\
\bottomrule
\end{tabular}
\end{table}

On this benchmark the documented covariance equals the true one, since the noise is generated with it, so documented WLS and oracle WLS are the same method and we report only oracle WLS. The de Chalendar heuristic makes the covariance implicit, assuming variances proportional to signal magnitudes,
\begin{equation} \label{eq:dechalendar-weights}
\boldsymbol{\Sigma}_{ii}(t) \;=\; \frac{\max\bigl(|\Xi_i(t)|, \gamma\bigr)}{\kappa},
\end{equation}
where $\Xi_i(t)$ is the rolling mean of signal $i$ and $\kappa, \gamma$ are constants of the heuristic. Their published objective minimizes the squared adjustments $\delta_i^2$, each divided by $\boldsymbol{\Sigma}_{ii}(t)$, over $\boldsymbol{\delta} = \mathbf{x} - \mathbf{y}_t$, which is the program~\eqref{eq:wls-qp} after substituting $\mathbf{x} = \mathbf{y}_t + \boldsymbol{\delta}$. We keep their preprocessing, in which static and rolling thresholds reject implausible values and rejected or missing cells are imputed by interpolation, and drop their aggregation and anti-symmetry constraints, since our state has no aggregate signals and represents each border by two non-negative directed flows instead of one signed flow.

The all-signal root-mean-square is about $10$~GW, so an overall RMSE near $100$~MW is a relative error of about one percent, though individual signals vary widely. The diffusion EB prior and the oracle WLS approximate the posterior with an ensemble of 64 draws each, and their point estimate is the ensemble mean, which for the oracle WLS coincides with the WLS solution. The Gaussian EB point estimate is its posterior mean, in closed form. The expected squared error of an individual draw exceeds that of the posterior mean by the posterior variance, so scoring draws would penalize the prior for honestly reporting its spread. RMSE then compares each method's best point summary. A posterior mean cannot represent a multimodal or wide posterior, so the distribution itself is judged separately, by the calibration and CRPS results of Appendix~\ref{app:additional-results}.

\section{Additional results}
\label{app:additional-results}

\subsection{Per-component accuracy}

\begin{table}[h]
\centering
\caption{RMSE (MW) on the six-zone system, all test hours. The generation, load, and flow columns report the per-component RMSE, and the overall column is the all-signal RMSE.}
\label{tab:base-results}
\renewcommand{\arraystretch}{1.15}
\begin{tabular}{lrrrr}
\toprule
Method & Gen & Load & Flow & Overall \\
\midrule
Raw observations & 131.7 & 151.5 & 200.0 & 146.4 \\
Naive WLS & 129.5 & 147.5 & 187.9 & 141.9 \\
Oracle WLS & 120.3 & 147.4 & 185.6 & 134.9 \\
Constrained WLS QP & 103.7 & 149.7 & 147.9 & 115.1 \\
De Chalendar & 237.5 & 729.9 & 233.5 & 287.3 \\
Gaussian EB prior & 106.2 & 141.0 & 153.8 & 117.6 \\
\textbf{Diffusion EB prior} & \textbf{92.5} & \textbf{135.0} & \textbf{117.5} & \textbf{99.8} \\
\bottomrule
\end{tabular}
\end{table}

Table~\ref{tab:base-results} reports RMSE per component on \textsc{eu6}. The non-negativity constraints lower the flow residual of the constrained WLS QP relative to the unconstrained naive and oracle WLS, 147.9~MW against roughly 186~MW, and the diffusion EB prior lowers it further to 117.5~MW while also attaining the lowest generation and load residuals.

The de Chalendar weights suit the EIA data they were built for, where reporting error grows with the size of the balancing authority~\cite{de2021physics}. The heuristic needs no noise covariance and no historical corpus, which makes it the reference for real-time reconciliation where neither is available.

\begin{figure*}[t]
\centering
\includegraphics[width=0.85\linewidth]{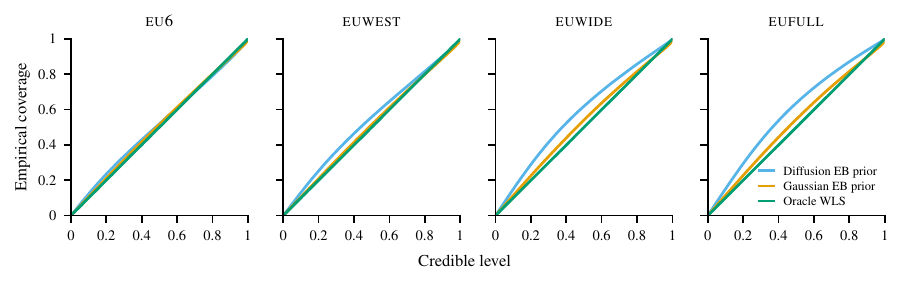}
\caption{Empirical coverage against credible level for the diffusion posterior, the Gaussian EB posterior, and the oracle WLS. The oracle WLS and the Gaussian EB lie on the diagonal, and the diffusion posterior turns conservative at intermediate levels on the larger systems.}
\label{fig:calibration}
\end{figure*}

\subsection{Prior fidelity} \label{ssec:prior-fidelity}

\begin{figure*}[t]
\centering
\begin{minipage}[c]{0.49\linewidth}\centering
\includegraphics[width=\linewidth]{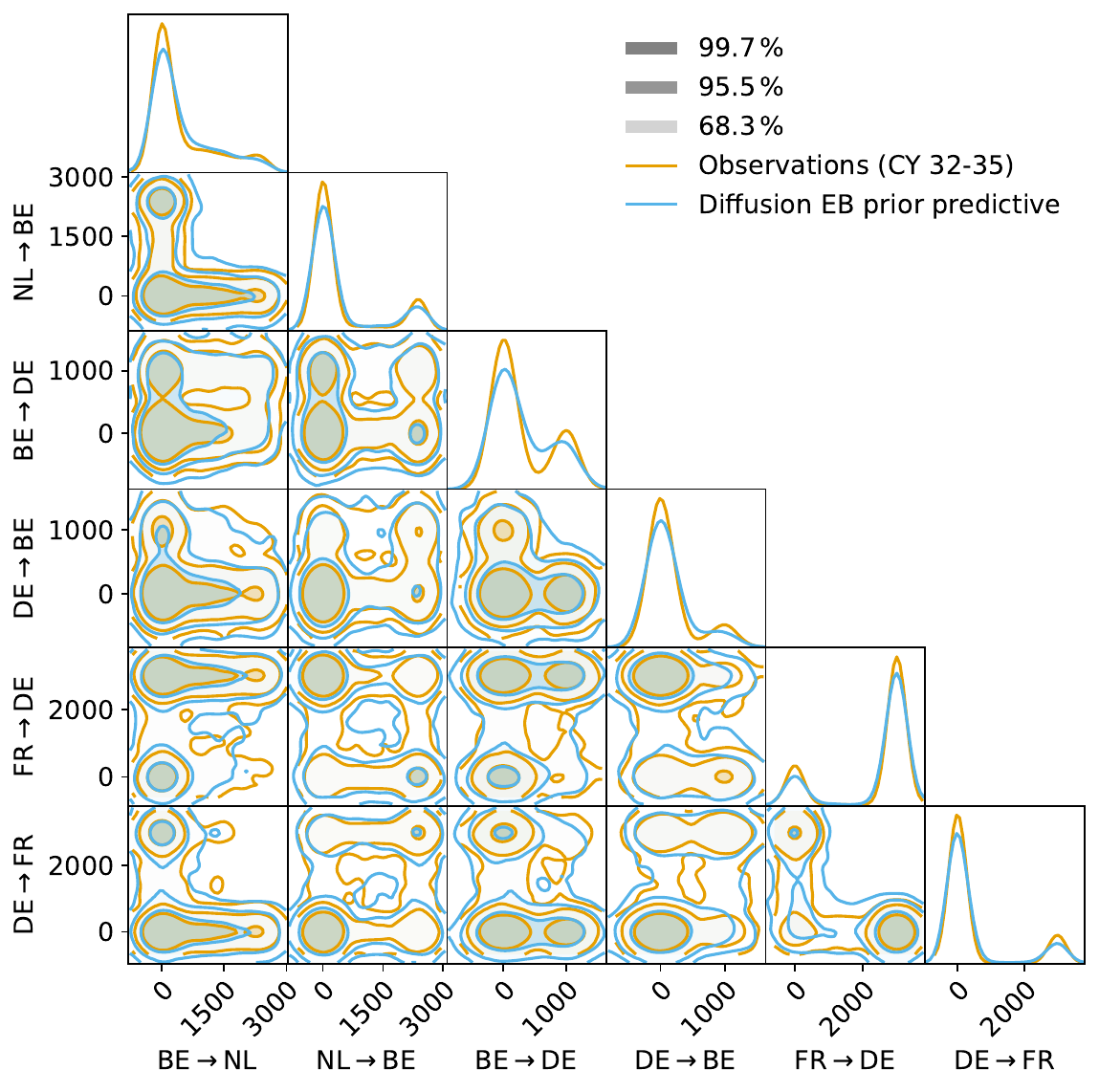}
\end{minipage}\hfil
\begin{minipage}[c]{0.49\linewidth}\centering
\includegraphics[width=\linewidth]{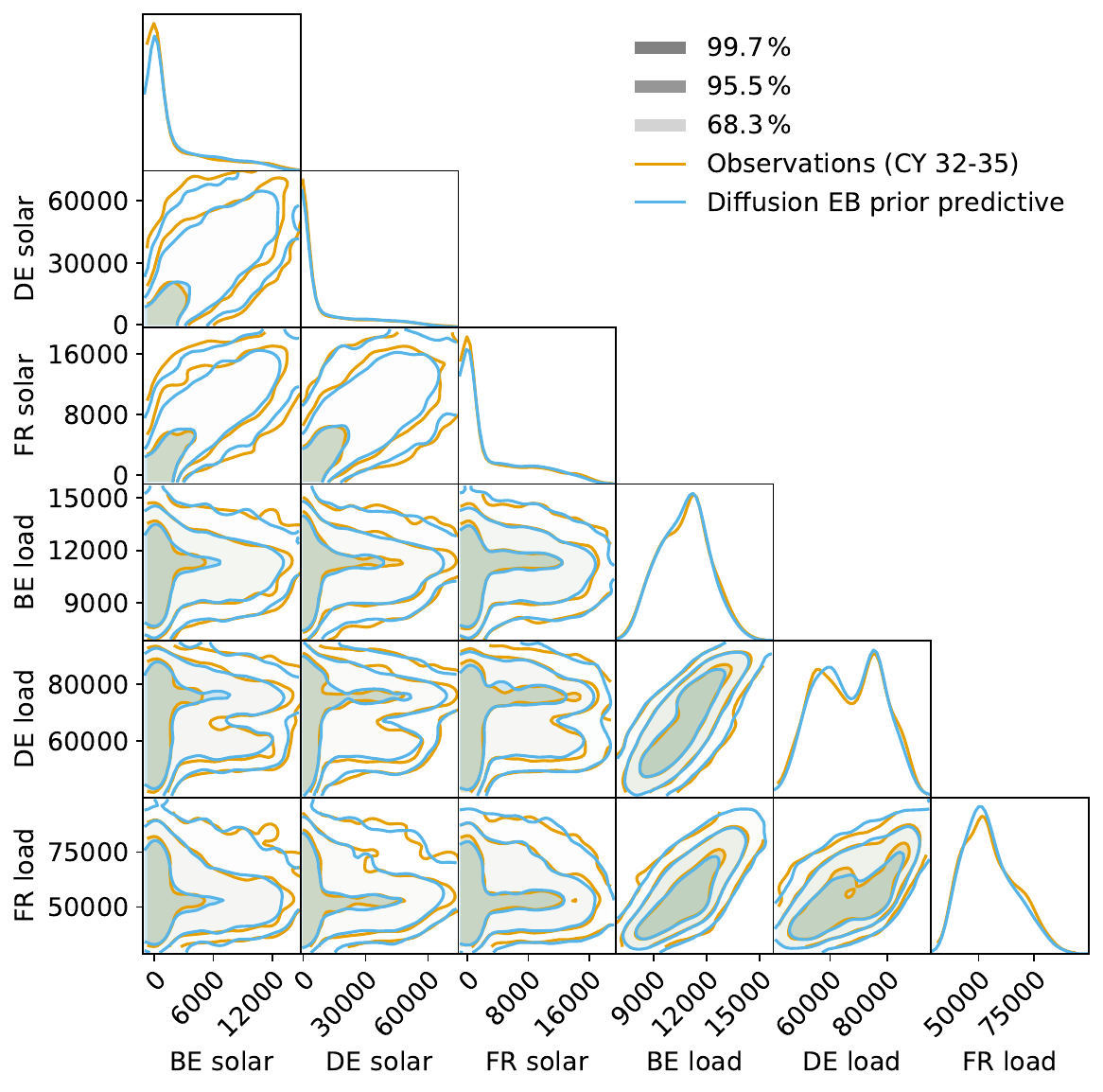}
\end{minipage}
\caption{Prior predictive check: unconditional prior samples plus the documented noise (blue) against the test set (orange). Left: the directed flows of the BE-NL, BE-DE, and FR-DE borders, each pair active one direction at a time. Right: solar and load in BE, DE, and FR.}
\label{fig:prior-corner}
\end{figure*}

Comparing prior samples directly to the test states needs the ground truth, so that check runs only on synthetic data. The same question has a ground-truth-free form usable on real data, the prior predictive check. We draw unconditional prior samples, add the documented noise to form the prior predictive $\mathbf{y}_{\mathrm{rep}} = \mathbf{x} + \boldsymbol{\varepsilon}$, and compare the two distributions with per-signal and sliced Wasserstein distances and the standard-deviation mismatch (Table~\ref{tab:prior-predictive}). On all four systems the diffusion EB prior predictive matches the observations closely, with per-signal standard deviations within 8 to 12 percent, and on \textsc{eu6} it is on par with the oracle. The balance-gap magnitude is reproduced to within about one percent on every system, for example a root-mean-square $\|\mathbf{C}\mathbf{y}_{\mathrm{rep}}\|$ of 705~MW against 709~MW in the observations on \textsc{eu6}. The excess width seen against the states is mostly hidden here, because the observation noise is present in both the predictive and the data, so an over-wide prior plus noise still lands near the observed spread. Figure~\ref{fig:prior-corner} shows the check on the directed flows of three borders and on solar and load in three zones. The state distribution is a mixture of operating regimes, a border exports in one direction or the other, a line runs at its transfer limit or idles, solar is zero at night, and load and solar rise together during the day across zones. The prior recovers the location, the weight, and the cross-signal coupling of these regimes from noisy observations alone.
\begin{table}[t]
\centering
\caption{Prior predictive check against the test set, in observation-standard-deviation units. Lower is closer.}
\label{tab:prior-predictive}
\renewcommand{\arraystretch}{1.15}
\begin{tabular}{llrrr}
\toprule
System & Prior & Marginal $W_1$ & Sliced-$W$ & Std mismatch \\
\midrule
\textsc{eu6}    & Oracle       & 0.05 & 0.06 & 0.07 \\
\textsc{eu6}    & Diffusion EB & 0.08 & 0.05 & 0.08 \\
\textsc{euwest} & Diffusion EB & 0.10 & 0.06 & 0.08 \\
\textsc{euwide} & Diffusion EB & 0.13 & 0.08 & 0.12 \\
\textsc{eufull} & Diffusion EB & 0.13 & 0.08 & 0.12 \\
\bottomrule
\end{tabular}
\end{table}

\subsection{Calibration} \label{ssec:calibration}

Calibration is tested as follows. At each level $\alpha$, the 1024 draws per time give one central credible interval per signal, and the empirical coverage is the fraction of these intervals that contain the state. The check uses the test states, not draws from the learned prior, so nominal coverage is not guaranteed by construction, and it tests the prior, the likelihood, and the sampler together, one signal at a time. Figure~\ref{fig:calibration} plots empirical coverage against $\alpha$ over the active signals, those whose state standard deviation over the test set exceeds 1~MW, because the structurally constant signals are covered trivially by every method and would only raise the coverage. A trustworthy posterior lies on the diagonal. The oracle WLS lies on the diagonal on every system, a control for the diagnostic, and the Gaussian EB stays close to nominal throughout. The diffusion posterior turns conservative at intermediate levels on the two larger systems, up to 14 points above the diagonal on \textsc{eufull}, and its 90\% coverage stays between 88.6\% and 93.5\% on all four.

Table~\ref{tab:calibration-systems} reports the mean 90\% interval width. The diffusion posterior is 31\% to 41\% below the oracle WLS and 15.6\% below the Gaussian EB at six zones, a gap that narrows to 2.0\% at thirty-five. The oracle WLS is calibrated only by being wide, and the Gaussian EB is calibrated on average but its posterior covariance is the same every hour and cannot adapt.

\begin{table}[h]
\centering
\caption{Mean 90\% interval width in MW over the active signals, full test set ($35\,040$ hours).}
\label{tab:calibration-systems}
\renewcommand{\arraystretch}{1.15}
\setlength{\tabcolsep}{4.5pt}\small
\begin{tabular}{lrrr}
\toprule
System & Diffusion EB & Gaussian EB & Oracle WLS \\
\midrule
\textsc{eu6}    & 291 & 344 & 419 \\
\textsc{euwest} & 308 & 333 & 442 \\
\textsc{euwide} & 345 & 303 & 472 \\
\textsc{eufull} & 347 & 284 & 471 \\
\bottomrule
\end{tabular}
\end{table}

\subsection{Robustness to noise misspecification} \label{ssec:robustness}

The documented covariance is a belief, and in deployment a wrong belief enters everywhere at once. The E-step removes the believed noise during training, the Gaussian EB subtracts it from the observation covariance, and the likelihood uses the same belief again at inference, while the data keep their true noise. We therefore re-estimate both learned priors with the believed covariance scaled by $s_{\mathrm{doc}} \in \{0.5, 2\}$ and evaluate on the unchanged test observations with the same wrong belief. We use \textsc{eu6}, where the diffusion training is cheapest. The constrained WLS QP receives the same belief, but its minimizer is unchanged by a global rescaling of the weights, so the small variation in its row is numerical and the stress falls on the learned priors. The experiment misspecifies only the overall scale of the belief, so every variance is off by the same factor. A belief that is wrong about which signals are noisy is not tested.

\begin{table}[h]
\centering
\caption{Noise-belief misspecification on \textsc{eu6}, full test set. Overall RMSE in MW under the believed covariance $s_{\mathrm{doc}} \boldsymbol{\Sigma}$.}
\label{tab:sigma-misspec}
\renewcommand{\arraystretch}{1.15}
\begin{tabular}{lrrr}
\toprule
 & $s_{\mathrm{doc}}{=}0.5$ & $s_{\mathrm{doc}}{=}1$ & $s_{\mathrm{doc}}{=}2$ \\
\midrule
Constrained WLS QP & 114.1 & 115.1 & 118.6 \\
Gaussian EB prior  & 121.6 & 117.6 & 151.2 \\
\textbf{Diffusion EB prior} & \textbf{112.9} & \textbf{99.8} & \textbf{104.9} \\
\midrule
Diffusion EB gap to QP  & $1.1\%$ & $13.3\%$ & $11.6\%$ \\
Diffusion EB gap to Gaussian EB & $7.2\%$ & $15.1\%$ & $30.6\%$ \\
\bottomrule
\end{tabular}
\end{table}

Table~\ref{tab:sigma-misspec} shows the two learned priors failing in opposite directions while the diffusion prior stays ahead everywhere. For the diffusion prior the gap to the constrained WLS QP is $11.6\%$ at $s_{\mathrm{doc}} = 2$ against $13.3\%$ under the correct belief, and understating leaves 112.9 against 114.1~MW at $s_{\mathrm{doc}} = 0.5$, because the E-step and the inference likelihood then both over-trust the noisy observations. A plausible reason for the robustness of the diffusion prior is that the EM loop drives it toward consistency of prior plus believed noise with the observations, so the prior itself absorbs part of the wrong belief. The misspecified scale is also detectable, since the balance residual $\mathbf{C}\mathbf{y}_t$ has covariance $\mathbf{C}\boldsymbol{\Sigma}\mathbf{C}^\top$, so a wrong scale appears as residuals larger or smaller than the belief predicts and can be corrected before training.

\end{document}